\documentclass[12pt]{article}
\usepackage{amsmath}
\usepackage{graphicx,authblk}
\usepackage{enumerate,color,rotating,booktabs,multirow,changepage,anyfontsize}
\usepackage{url} 

\newcommand{\blind}{0}

\begin{document}

\def\spacingset#1{\renewcommand{\baselinestretch}%
{#1}\small\normalsize} \spacingset{1}


\if0\blind
{
  \title{\bf Outcome measurement error correction for survival analyses with multiple failure types: application to hearing loss studies}
  \author{Yujie Wu\\
    \textit{Department of Biostatistics, Harvard University}\\
    and \\
    Molin Wang \\
    \textit{Departments of Biostatistics and Epidemiology, Harvard University,}\\
\textit{Channing Division of Network Medicine, Brigham and Women's Hospital,}\\\textit{and Harvard Medical School}}
\date{}
  \maketitle
} \fi

\bigskip
\begin{abstract}
In epidemiological studies, participants' disease status is often collected through self-reported outcomes in place of formal medical tests due to budget constraints. However, self-reported outcomes are often subject to measurement errors, and may lead to biased estimates if used in statistical analyses. In this paper, we propose statistical methods to correct for outcome measurement errors in survival analyses with multiple failure types through a reweighting strategy. We also discuss asymptotic properties of the proposed estimators and derive their asymptotic variances. The work is motivated by Conservation of Hearing Study (CHEARS) which aims to evaluate risk factors for hearing loss in the Nurses' Health Studies II (NHS II). We apply the proposed method to adjust for the measurement errors in self-reported hearing outcomes; the analysis results suggest that tinnitus is positively associated with moderate hearing loss at both low or mid and high sound frequencies, while the effects between different frequencies are similar.
\end{abstract}

\noindent%
{\it Keywords:} Outcome measurement error; Misclassification of failure types; {\color{black}{Multiple failure types; Correlated survival times}}; Weighted estimating equations
\vfill

\newpage
\spacingset{1.9} 

\section{Introduction}

Hearing loss is a  leading cause of disability; more than 5\% of the world's population suffers from disabling hearing loss \cite{WHO, wilson2017global}. Since hearing loss is usually irreversible, research to identify risk factors for hearing loss can be beneficial for early intervention and treatment \cite{curhan2019world}. Hearing loss is diagnosed through audiometric hearing tests where, for each ear the lowest decibel (dB) level tone that can be heard is recorded as the hearing threshold at a given sound frequency. The hearing acuity is then categorized based on pure-tone average (PTA) of the thresholds for low-frequency (0.5, 1, 2kHz), mid-frequency (3, 4 kHz) and high-frequency (6, 8 kHz) \cite{curhan2018adherence}, and moderate hearing loss is defined as the PTA being greater than 35 dB at the better hearing ear \cite{olusanya2019hearing}. However, in large epidemiological studies  for evaluating the association between a potential risk factor and incident hearing loss, it may be infeasible to enroll all study participants to take {\color{black}{medical}} tests due to time and budget constraints; therefore, the disease outcomes are often collected through questionnaire-based self-reporting, which maybe prone to measurement error.  In a validation study that compared self-reported hearing loss obtained among women less than 70 years old and true hearing loss ascertained by formal audiometric tests, the sensitivity of detecting true moderate or severe hearing loss based on self-reported hearing status is 95\% \cite{sindhusake2001validation}. However, the specificity was only 65\%, likely leading to biased association estimates. Another disadvantage of using self-reported hearing status as the outcome is that it prevents from understanding the etiology of hearing loss at different frequencies, since the questionnaires only asked people's overall hearing status without detailed information on their hearing abilities at different sound frequencies.

{\color{black}{To address the challenge of  mis-measured outcomes in statistical analyses, we propose statistical methods to correct for measurement error in outcomes in the setting of survival analyses with multiple event types.}} Our work is motivated by the Conservation of Hearing Study (CHEARS), which evaluates risk factors of hearing loss among participants in the Nurses’ Health Studies II (NHS II), an ongoing cohort study consisting of 116,430 female registered nurses in the United States, aged 25-42 years at enrollment in 1989 \cite{curhan2018adherence}. In NHS II, participants filled in questionnaires for lifestyle information every two years. Self-reported hearing status was classified into four categories: `no', `mild', `moderate' and `severe', and were asked in the 2009, 2013 and 2017 questionnaires. Participants who reported hearing problems additionally provided the age at which they noticed a change in their hearing status. Due to its irreversibility, researchers often focus on the time to first self-reported hearing loss and apply survival analysis methods to investigate its risk factors \cite{lin2020cigarette, gupta2020predictive, gupta2019type}. Typically, moderate or worse hearing loss is used as the primary outcome of interest due to its clinical meaningfulness and minimization of misclassification \cite{kamil2015factors, ferrite2011validity, schow1990self, gomez2001comparison, sindhusake2001validation}. The CHEARS Audiology Assessment Arm (AAA)  is a sub-cohort of NHS II; the hearing thresholds of AAA participants were collected through audiometric tests at 19 geographically diverse testing sites. The baseline testing was conducted on 3,749 participants and 3-year follow-up testing was completed on 3,136 participants (84\%) in June 2018 \cite{curhan2020prospective}. The availability of both the true audiometric hearing thresholds, which can be used to derive the gold standard hearing status, and the self-reported hearing status in AAA enables us to use it as a validation study to correct for the error-prone self-reported hearing outcomes in CHEARS. The gold standard hearing loss outcome has multiple subtypes, hearing loss under low-, mid- and high-frequencies {\color{black}{and the occurrence of hearing loss at one sound frequency will not preclude the occurrence of hearing loss at other sound frequencies}}. In this paper, we will develop statistical methods to correct for outcome measurement error in survival analysis with multiple event types under the framework of Wei, Lin, and Weissfeld (WLW) marginal model \cite{wei1989regression}.

The WLW model was proposed to analyze data of multiple types of failure. It models the marginal hazard of type $j$ failure time through a Cox regression model without assuming any particular structure of dependence among different failure times. The coefficients can be estimated by maximizing the event-specific partial likelihood and the sandwich variance estimator is used to estimate the covariance matrix of the estimated coefficients \cite{wei1989regression}. To the best of our knowledge, no method has been proposed to correct for outcome measurement error in survival analyses  with multiple failure types. A string of methods are available for Cox regression with missing failure status where they adopt a reweighted Cox regression approach with weights being the probability of the event occurring at a particular time \cite{snapinn1998survival,cook2000adjusting,cook2004analysis}. Our method is also a weight-based method, and it differs from these existing methods mainly in the following aspects. First, we consider the complex setting of multiple event types while the existing methods are for the single event type settings. {\color{black}{Second, when deriving the weights, the existing methods assume that the occurrence of the true outcome at different potential event times is independent between each other which is often violated in real practice, while our method does not require this assumption. }} 

The remainder of the paper is organized as follows. In section 2, we introduce our proposed methods. In Section 3, we conduct extensive simulation studies to evaluate the finite sample performance of our proposed methods. In Section 4, we apply our method to CHEARS to investigate the effect of tinnitus on hearing acuity at different frequencies. We provide a concluding discussion in Section 5.

\section{Methods}
\subsection{WLW method}
The WLW method models the marginal hazard of each failure outcome through a Cox proportional hazards model without assuming any dependence structure between different failure times. For the $k$-th event ($k=1,\ldots,K$), let $T_{ki}$ be the failure time for the $i$-th individual $(i=1,\ldots,n)$, and $C_{ki}$ be the corresponding censoring time. Let $X_{ki}=\min(T_{ki}, C_{ki})$ and {\color{black}{$\Delta_{ki}=I(T_{ki}\le C_{ki})$}}, where $I()$ is the indicator function. We define the counting process of event $k$ for the $i$-th individual as: $N_{ki}(t)=I(X_{ki}\le t, \Delta_{ki}=1)$ and the `at risk' indicator as: $Y_{ki}(t)=I(X_{ki}\ge t)$. The hazard function for the $k$-th event is defined as:
\begin{equation*}
	\lambda_{ki}(t)=\lambda_{k0}(t)\exp(\boldsymbol{\beta}_k^T\boldsymbol{Z}_{ki}(t))
\end{equation*}
where $\lambda_{k0}(t)$ is the event-specific baseline hazard function, $\boldsymbol{Z}_{ki}(t)$ is a column vector containing possible time-dependent covariates for the $k$-th event type, {\color{black}{and we assume that $C_{ki}$ is independent of $T_{ki}$ given $\boldsymbol{Z}_{ki}(t)$.}} In the WLW method, parameter estimates are obtained by solving the following event-specific estimating equations \cite{andersen1982cox}:
\begin{equation}
	\boldsymbol{U}_k(\boldsymbol{\beta}_k)=\sum_{i=1}^{n}\int_{0}^{\infty}\left\{ \boldsymbol{Z}_{ki}(t)-\frac{ \sum_{j=1}^{n}\boldsymbol{Z}_{kj}(t)Y_{kj}(t)\exp(\boldsymbol{\beta}_k^T\boldsymbol{Z}_{kj}(t))   }{\sum_{j=1}^{n}Y_{kj}(t)\exp(\boldsymbol{\beta}_k^T\boldsymbol{Z}_{kj}(t)) }  \right\}\,dN_{ki}(t)=\boldsymbol{0},
	\label{WLW_est}
\end{equation}

A sandwich variance estimator is used to estimate the variance-covariance matrix of the estimated parameters $\widehat{\boldsymbol{\beta}}=(\widehat{\boldsymbol{\beta}}_1^T,\ldots,\widehat{\boldsymbol{\beta}}_K^T)^T$, and we refer readers to \cite{wei1989regression} for details. Note that $\boldsymbol{Z}_{ki}(t)$ are event-specific covariates, and are not necessarily the same across different event types. If there are shared covariates across event types and is believed to have a common effect on the outcomes, for instance, the first element $\beta_{k,1}$ in each $\boldsymbol{\beta}_k$, representing the treatment effect of the same exposure, an inverse variance pooled estimator was proposed to estimate the common effect $\eta$ through the following form:
\begin{equation*}
	\widehat{\eta}=\sum_{k=1}^{K}c_k\widehat{\beta}_{k,1}
\end{equation*}
where $(c_1,\ldots,c_K)^T=\left(\boldsymbol{e}^T\widehat{\boldsymbol\Sigma}\boldsymbol{e}\right)^{-1}\widehat{\boldsymbol\Sigma}\boldsymbol{e}$,  $\boldsymbol{e}$ is a length-K column vector with element 1 and $\widehat{\boldsymbol{\Sigma}}$ is the estimated variance-covariance matrix of $(\widehat{\beta}_{1,1}, \widehat{\beta}_{2,1},\dots, \widehat{\beta}_{K,1})^T$ \cite{wei1989regression}.


\subsection{Outcome measurement error-corrected WLW model under the MS/EVS study design}
\subsubsection{Estimating equations}
In large epidemiological studies, true disease statuses may not be available due to budget constraint, and self-reported outcomes, often collected through questionnaires, are available instead. {\color{black}{Let $\mathcal{T}_{ki}$ be the collection of possible event times {\color{black}{(e.g. questionnaire return times)}} of the $k$-th event for the $i$-th individual.}} We denote the self-reported event time for the $k$-th event as $T_{ki}^\ast$ and the self-reported event status at questionnaire time $t$ as $\delta^\ast_{ki}(t)$, where $\delta^\ast_{ki}(t)=1$ if participants have the $k$-th event at time $t$ and 0 otherwise. {\color{black}{Since we focus on incidence disease outcome in this paper, once participants report the event at some time $t$, they will always have the event afterwards (i.e. $\delta_{ki}^\ast(t')=1$ for $t'\ge t$, if $\delta_{ki}^\ast(t)=1$).}} Moreover, we use $\delta_{ki}(t)$ to denote the true event status for the $k$-th event at questionnaire time $t$, where $\delta_{ki}(t)=1$ represents having the true $k$-th event {\color{black}{at}} time $t$ and 0 otherwise. Similar to $\delta_{ki}^\ast(t)$, we also assume $\delta_{ki}(t') = 1$ for $t' \ge t,$ if $\delta_{ki}(t)=1$. Since the self-reported events are subject to measurement error, directly using them in analyses may lead to biased estimate of regression parameters. In this section, we propose statistical methods to correct for outcome measurement error-caused bias in the WLW model.

{\color{black}{In the main study, only the self-reported event time and status $\{T^\ast_{ki}, \delta^\ast_{ki}(t); t\in\mathcal{T}_{ki}, k=1,\ldots,K, i=1,\ldots,n_M\}$ are available, where $n_M$ denotes the sample size of the main study. In the validation study, both the true and self-reported event time and status $\{T_{ki}, \delta_{ki}(t), T^\ast_{ki}, \delta^\ast_{ki}(t); t\in\mathcal{T}_{ki}, k=1,\ldots,K, i=1,\ldots,n_V\}$ are available, where $n_V$ denotes the sample size of the validation study. All covariates of interests, which are free of measurement error, are available in both the main and validation studies.}} In this paper, we consider both the main study/external validation study (MS/EVS) design and the main study/internal validation study (MS/IVS) design. A validation study is \textit{internal} if participants in the validation study are representative {subsample} of the main study; otherwise the validation study is external \cite{spiegelman2001efficient}. When considering the MS/EVS design, we assume transportability, which requires that the main and validation studies share the same outcome measurement error process. Under transportability, the measurement error model fitted in the validation study can be used to calibrate the mismeasured outcomes in the main study \cite{carroll2006measurement, ackerman2019transportability}. We focus on the methods under MS/EVS design in this section. The methods under MS/IVS design are discussed in the next section. 

In epidemiological studies, self-reported outcomes are usually obtained at each questionnaire period, and researchers use time to the first self-reported event as the event time for incidence analyses. For example, in CHEARS, the hearing status is collected through a biannually distributed questionnaire return. We therefore introduce a counting process $N^\ast_{ki}(t)$ that jumps with size one at each follow-up/questionnaire return time. {\color{black}{Since the true event time and status are unavailable for participants in the main study, we consider each questionnaire time as a potential event time, and therefore $N^\ast_{ki}(t)$ jumps at each questionnaire time regardless of whether the $k$-th event is reported by the participants or not. }} 
 It can be shown that when the true outcome is available, the counting process for the true $k$-th event type is a function of the counting process of the potential event times \cite{cook2004analysis}:	
\begin{equation}
	N_{ki}(t)=1-\prod_{s=0}^{t}[1-\delta_{ki}(s)dN^\ast_{ki}(s)]
	\label{counting_process}
\end{equation}

When the true outcomes are not available, we propose to replace $\delta_{ki}(s)$ in (\ref{counting_process}) with its corresponding conditional expectation:
\begin{equation}
	\begin{split}
		\widetilde{N}_{ki}(t)&=1-E\left(\prod_{s=0}^{t}[1-\delta_{ki}(s)dN_{ki}(s)]|\mathcal{F}_t\right)\\
		&=1-\prod_{s=0}^{t}[1-h(\boldsymbol{\gamma}_k, \boldsymbol{W}_{ki}(s))dN_{ki}(s)],
	\end{split}
	\label{N_tilde}
\end{equation}
where $\mathcal{F}_t$ contains the history of all possible time-dependent covariates $\boldsymbol{W}_{ki}(t)$, that are predictive of the true $k$-th event status, such as time, self-reported outcomes and exposures etc., and these covariates can be determined by subject matter knowledge. The second equation in (\ref{N_tilde}) holds by applying law of iterated expectation, where $h(\boldsymbol{\gamma}_k, \boldsymbol{W}_{ki}(t))$ represents the conditional probability of the true $k$-th event occurring at the $t$-th questionnaire return time, given that it is $k$-th event-free up until the $t-1$-th questionnaire return time. Therefore,  $\widetilde{N}_{ki}(t)$ represents the probability of the true $k$-th event being observed by time $t$, given $\{N_{ki}(s), \boldsymbol{W}_{ki}(s), 0\le s\le t\}$. Furthermore, the `at risk' indicator can be modified as $
\widetilde{Y}_{ki}(t)=[1-\widetilde{N}_{ki}(t-)]I(C_{ki}\ge t)
$, and it represents the probability that the true $k$-th event occurs at some time $t'\ge t$, given $\{N_{ki}(s), \boldsymbol{W}_{ki}(s), 0\le s\le t\}$.

The conditional probabilities $h(\boldsymbol{\gamma}_k, \boldsymbol{W}_{ki}(t))$ can be estimated through fitting measurement error models in the external validation set where the true event status $\delta_{ki}(t)$ are available. The interpretation of $h(\boldsymbol{\gamma}_k, \boldsymbol{W}_{ki}(t))$, as described above, motivates us to use the pooled logistic regression as the measurement error models \cite{d1990relation}. The pooled logistic regression proceeds by dividing a person's entire follow-up time into several sub-intervals, and it models the probability of the event occurring at the end of the sub-interval provided that they are event-free at the start of the sub-interval:
\begin{equation*}
	\text{logit }P\left[\delta_{ki}(t)=1|\delta_{ki}(t-1)=0,\boldsymbol{W}_{ki}(t)\right]=\boldsymbol{\gamma}_k^T\boldsymbol{W}_{ki}(t).
\end{equation*}
Since we need to fit $K$ event-specific measurement error models, to account for possible correlations between the multiple outcomes, {\color{black}{we propose to use joint unbiased estimating functions by stacking the $K$ score functions for estimating $(\boldsymbol{\gamma}_1, \ldots, \boldsymbol{\gamma}_K)$, respectively. The score function for estimating $\boldsymbol{\gamma}_k$ is:
\begin{equation*}
	\begin{split}
		\boldsymbol{U}_{k}(\boldsymbol{\gamma}_k)&=\frac{1}{n_V}\sum_{i=1}^{n_V}\pi_{ki}(\boldsymbol{\gamma}_k)\\
		&=\frac{1}{n_V}\sum_{i=1}^{n_V}\sum_{t\in\mathcal{T}_{ki}}\left(\Big[\delta_{ki}(t)-\frac{1}{1+\exp(-\boldsymbol{\gamma}_k^T\boldsymbol{W}_{ki}(t))}\Big]\boldsymbol{W}_{ki}(t)\right).
	\end{split}
\end{equation*}		
}} For point estimates, it is equivalent to fitting ordinary pooled logistic regressions for each event type separately, and variance can be estimated using the sandwich formula.

When the true outcomes are not available, we can replace $N_{ki}(t)$ and $Y_{ki}(t)$ in Equation (\ref{WLW_est}) with $\widetilde{N}_{ki}(t)$ and $\widetilde{Y}_{ki}(t)$, leading to the following weighted WLW estimating equations:

\begin{equation}
	\scriptsize
	\begin{split}
		\widetilde{\boldsymbol{U}}_k(\boldsymbol{\beta}_k, \boldsymbol{\gamma}_k)=&\frac{1}{n_M}\sum_{i=1}^{n_M}\phi^E_{ki}(\boldsymbol{\beta}_k,\boldsymbol{\gamma}_k)\\
		=&\frac{1}{n_M}\sum_{i=1}^{n_M}\int_{0}^{\infty}\left\{ \boldsymbol{Z}_{ki}(t)-\frac{ \sum_{j=1}^{n_M}\boldsymbol{Z}_{kj}(t)\widetilde{Y}_{kj}(t)\exp(\boldsymbol{\beta}_k^T\boldsymbol{Z}_{kj}(t))   }{\sum_{j=1}^{n_M}\widetilde{Y}_{kj}(t)\exp(\boldsymbol{\beta}_k^T\boldsymbol{Z}_{kj}(t)) }  \right\}\,d\widetilde{N}_{ki}(t)\\
		=&\frac{1}{n_M}\sum_{i=1}^{n_M}\sum_{t\in\mathcal{T}_{ki}}\left(  h(\boldsymbol{W}_{ki}(t);\boldsymbol{\gamma_k})\prod_{t'<t, t'\in\widetilde\mathcal{T}_{ki}}  (1-h(\boldsymbol{W}_{ki}(t');\boldsymbol{\gamma}_k)) \right)\\
		&\times\left( \boldsymbol{Z}_{ki}(t)-\frac{\sum_{j=1}^{n_M}\left( \prod_{t''<t, t''\in\widetilde\mathcal{T}_{kj}}[1-h(\boldsymbol{W}_{kj}(t'');\boldsymbol{\gamma}_k)] \right)I(C_{kj}\ge t)\boldsymbol{Z}_{kj}(t)\exp(\boldsymbol{\beta}_k^T\boldsymbol{Z}_{kj}(t))       }{   \sum_{j=1}^{n_M}\left( \prod_{t''<t, t''\in\widetilde\mathcal{T}_{kj}}[1-h(\boldsymbol{W}_{kj}(t'');\boldsymbol{\gamma}_k)] \right)I(C_{kj}\ge t)\exp(\boldsymbol{\beta}_k^T\boldsymbol{Z}_{kj}(t))       }       \right)\\
		=&\boldsymbol{0},
	\end{split}
	\label{brewslow_est_eq}
\end{equation}
where $\widetilde\mathcal{T}_{ki}$ further includes time origin {\color{black}{($t=0$)}}. We define $h(\boldsymbol{W}_{ki}(0), \boldsymbol{\gamma}_k)=0$ since in incidence analyses, all participants are free of the events of interest at study entry. Note that,  $h(\boldsymbol{W}_{ki}(t);\boldsymbol{\gamma_k})\prod_{t'<t, t'\in\widetilde\mathcal{T}_{ki}}  (1-h(\boldsymbol{W}_{ki}(t');\boldsymbol{\gamma}_k))$ represents the probability of the $k$-th event occurring at time $t$. Therefore, the intuition behind estimating equation (\ref{brewslow_est_eq}) is that since we do not know the exact onset time of the true $k$-th event for the $i$-th individual, each questionnaire return time should be a \textit{potential} event time and contributes to the estimating equations. However, their contributions are not treated equal, and are {\color{black}{weighted}} by the probability of the event occurring at each time, and the weighting probabilities reflect how likely the true $k$-th event will occur at a particular potential event time.

Since the measurement error model is fitted among an external validation study that are independent of the main study, the point estimates of $\boldsymbol{\beta}_k$ can be obtained in a two-step approach. In the first step, we fit the event-specific measurement error models in the validation study, and obtain $\widehat{\boldsymbol{\gamma}}_k$. In the second step, we obtain point estimates $\boldsymbol{\widehat{\beta}}^E_k$ by solving $\widetilde{\boldsymbol{U}}_k(\boldsymbol{\beta}_k; \boldsymbol{\gamma}_k)$, where ${\boldsymbol{\gamma}}_k$ is replaced by $\widehat{\boldsymbol{\gamma}}_k$, and we use superscript $E$ to indicate the MS/EVS design.

\subsubsection{Asymptotic property}

To account for the uncertainty from estimating ${\boldsymbol{\gamma}}_k$ from the validation study, we consider using the sandwich variance estimator to estimate the variance-covariance matrix of the estimated regression parameters. The joint estimating equations for both the parameters, $\boldsymbol{\beta}$, in the weighted WLW model fitted in the main study and the parameters, $\boldsymbol{\gamma}$, in the calibration models fitted in the validation study is:
{\color{black}{\begin{equation}
			[\widetilde{\boldsymbol{U}}_1(\boldsymbol{\beta}_1, \boldsymbol{\gamma}_1), \ldots, \widetilde{\boldsymbol{U}}_K(\boldsymbol{\beta}_K, \boldsymbol{\gamma}_K), \boldsymbol{U}_{1}(\boldsymbol{\gamma}_1),\ldots, \boldsymbol{U}_{K}(\boldsymbol{\gamma}_K)]=\boldsymbol{0}
			\label{full_wlw_est}
		\end{equation}
}}

Under mild conditions, the solutions $\widehat{\boldsymbol{\beta}}^E_k$ to Equations (\ref{full_wlw_est}) converge in probability to the true ${\boldsymbol{\beta}}_k$ \cite{cook2004analysis}, and we provide an outline of the proof in the supplementary material section 1.1. Moreover, asymptotically, the collection of parameters across event types $\widehat{\boldsymbol{\beta}}^E$ and $\widehat{\boldsymbol{\gamma}}$ are normally distributed as
\begin{equation}
	\begin{bmatrix}
		\sqrt{n_M}(\widehat{\boldsymbol{\beta}}^E-\boldsymbol{\beta})\\
		\sqrt{n_M}(\widehat{\boldsymbol{\gamma}}-\boldsymbol{\gamma})\\
	\end{bmatrix}\overset{\mathcal{L}}{\to}\text{MVN}\left(\boldsymbol{0}, (\boldsymbol{A}^E(\boldsymbol{\beta},\boldsymbol{\gamma}))^{-1}\boldsymbol{B}^E(\boldsymbol{\beta},\boldsymbol{\gamma})((\boldsymbol{A}^E(\boldsymbol{\beta},\boldsymbol{\gamma}))^{-1})^T\right),
	\label{normality}
\end{equation}
where $\boldsymbol{A}^E(\boldsymbol{\beta},\boldsymbol{\gamma})=\begin{bmatrix}
	A^E_{\boldsymbol{\beta}\boldsymbol{\beta}}(\boldsymbol{\beta},\boldsymbol{\gamma})&A^E_{\boldsymbol{\beta}\boldsymbol{\gamma}}(\boldsymbol{\beta},\boldsymbol{\gamma})\\
	\boldsymbol{0}&A^E_{\boldsymbol{\gamma}\boldsymbol{\gamma}}(\boldsymbol{\beta},\boldsymbol{\gamma})\\
\end{bmatrix}$,  $\boldsymbol{B}^E(\boldsymbol{\beta},\boldsymbol{\gamma})=\begin{bmatrix}
	\boldsymbol{B}^E_{\boldsymbol{\beta}}(\boldsymbol{\beta},\boldsymbol{\gamma})&\boldsymbol{0}\\
	\boldsymbol{0}&{\color{black}{\frac{1}{\rho}}}\boldsymbol{B}^E_{\boldsymbol{\gamma}}(\boldsymbol{\beta},\boldsymbol{\gamma})
\end{bmatrix}$, {\color{black}{ and $\rho$ is the ratio of $n_V$ to $n_M$}}. The proof for asymptotic  normality is provided in the supplementary material section 1.2. The elements $A^E_{\boldsymbol{\beta}\boldsymbol{\beta}}(\boldsymbol{\beta},\boldsymbol{\gamma})$, $A^E_{\boldsymbol{\beta}\boldsymbol{\gamma}}(\boldsymbol{\beta},\boldsymbol{\gamma})$, and $A^E_{\boldsymbol{\gamma}\boldsymbol{\gamma}}(\boldsymbol{\beta},\boldsymbol{\gamma})$ in  $\boldsymbol{A}^E(\boldsymbol{\beta},\boldsymbol{\gamma})$ are diagonal  block matrices, and the $k$-th block can be consistently estimated by: 
$\widehat{\boldsymbol{A}}^E_{\boldsymbol{\widehat\beta}_k\boldsymbol{\widehat\beta}_k}(\boldsymbol{\beta}_k,\boldsymbol{\gamma}_k)=\frac{1}{n_M}\sum_{i=1}^{n_M}\frac{\partial}{\partial\boldsymbol{\beta}_k}\phi^E_{ki}(\boldsymbol{\widehat\beta}_k,\boldsymbol{\widehat\gamma}_k)$, $\widehat{\boldsymbol{A}}^E_{\boldsymbol{\widehat\beta}_k\boldsymbol{\widehat\gamma}_k}(\boldsymbol{\beta}_k,\boldsymbol{\gamma}_k)=\frac{1}{n_M}\sum_{i=1}^{n_M}\frac{\partial}{\partial\boldsymbol{\gamma}_k}\phi^E_{ki}(\boldsymbol{\widehat\beta}_k,\boldsymbol{\widehat\gamma}_k)$ and $\widehat{\boldsymbol{A}}^E_{\boldsymbol{\widehat\gamma}_k\boldsymbol{\widehat\gamma}_k}(\boldsymbol{\beta}_k,\boldsymbol{\gamma}_k)=\frac{1}{n_V}\sum_{l=1}^{n_V}\frac{\partial}{\partial\boldsymbol{\gamma}_k}\pi_{kl}(\boldsymbol{\widehat\gamma}_k)$, respectively. Detailed formula can be found in the supplementary material section 1.2.  

Elements $\boldsymbol{B}^E_{\boldsymbol{\beta}}(\boldsymbol{\beta},\boldsymbol{\gamma})$ and $\boldsymbol{B}^E_{\boldsymbol{\gamma}}(\boldsymbol{\beta},\boldsymbol{\gamma})$ in $\boldsymbol{B}^E(\boldsymbol{\beta},\boldsymbol{\gamma})$ can also be partitioned into block matrices. The $(k,k')$-th block ($k,k'=1,\ldots,K$) of $\boldsymbol{B}^E_{\boldsymbol{\gamma}}(\boldsymbol{\beta},\boldsymbol{\gamma})$ can be estimated by $\widehat{\boldsymbol{B}}^E_{\boldsymbol{\widehat\gamma}_k,\boldsymbol{\widehat\gamma}_{k'} }(\boldsymbol{\beta},\boldsymbol{\gamma})=\frac{1}{n_V}\sum_{l=1}^{n_V}\pi_{kl}(\boldsymbol{\widehat\gamma}_k)\pi_{k'l}^T(\boldsymbol{\widehat\gamma}_{k'})$. The $(k,k')$-th block of $\boldsymbol{B}^E_{\boldsymbol{\beta}}(\boldsymbol{\beta},\boldsymbol{\gamma})$ can be estimated by:
\begin{equation*}
	\begin{split}
		\widehat{\boldsymbol{B}}^E_{\widehat{\boldsymbol{\beta}}_k,
			\widehat{\boldsymbol{\beta}}_{k'}}&=\frac{1}{n_M}\sum_{i=1}^{n_M}\widehat{\widetilde\phi}_{ki}^{E,\ast}(\widehat{\boldsymbol{\beta}}_k,\widehat{\boldsymbol{\gamma}}_k)[\widehat{\widetilde\phi}_{k'i}^{E,\ast}(\widehat{\boldsymbol{\beta}}_{k'},\widehat{\boldsymbol{\gamma}}_{k'})]^T,\quad k,k'=1,2\ldots,K,
	\end{split}
\end{equation*}

where \begin{equation*}
	\begin{split}
		\widehat{\widetilde\phi}_{ki}^{E,\ast}({\boldsymbol{\beta}}_k,{\boldsymbol{\gamma}}_k)&=\sum_{t\in\mathcal{T}_{ki}}\left(  h(\boldsymbol{W}_{ki}(t);\boldsymbol{\gamma_k})\prod_{t'<t, t'\in\widetilde\mathcal{T}_{ki}}  (1-h(\boldsymbol{W}_{ki}(t');\boldsymbol{\gamma}_k)) \right)\times\left( \boldsymbol{Z}_{ki}(t)-\frac{  \widetilde{S}^{(1)}_k(t;\boldsymbol{\beta}_k,\boldsymbol{\gamma}_k)  }{   \widetilde{S}^{(0)}_k(t;\boldsymbol{\beta}_k,\boldsymbol{\gamma}_k) }       \right)\\
		&-\sum_{j=1}^{n_M}\sum_{t\in\mathcal{T}_{kj}}\Bigg\{\left[h(\boldsymbol{W}_{kj}(t);\boldsymbol{\gamma_k})\prod_{t'<t, t'\in\widetilde\mathcal{T}_{kj}}  (1-h(\boldsymbol{W}_{kj}(t');\boldsymbol{\gamma}_k)) \right]\times\\
		&\left(  \frac{  \left(\prod_{t_{ki}< t, t_{ki}
				\in\widetilde\mathcal{T}_{ki}}(1-h(t_{ki};\boldsymbol{\gamma}_k))\right)I(C_{ki}\ge t)\exp(\boldsymbol{\beta}_k^T\boldsymbol{Z}_{ki}(t))  }{\widetilde{S}_k^{(0)}(t;\boldsymbol{\beta}_k,\boldsymbol{\gamma}_k)}\left[\boldsymbol{Z}_{ki}(t)-\frac{\widetilde{S}^{(1)}_k(t;\boldsymbol{\beta}_k,\boldsymbol{\gamma}_k)}{\widetilde{S}^{(0)}_k(t;\boldsymbol{\beta}_k,\boldsymbol{\gamma}_k)}\right]  \right)\Bigg\},
	\end{split}
\end{equation*}
$\widetilde{S}^{(0)}_k(t;\boldsymbol{\beta}_k,\boldsymbol{\gamma}_k)= \sum_{j=1}^{n_M}\left( \prod_{t'<t, t'\in\widetilde\mathcal{T}_{kj}}[1-h(\boldsymbol{W}_{kj}(t');\boldsymbol{\gamma}_k)] \right)I(C_{kj}\ge t)\exp(\boldsymbol{\beta}_k^T\boldsymbol{Z}_{kj}(t))      $, and $\widetilde{S}^{(1)}_k(t;\boldsymbol{\beta}_k,\boldsymbol{\gamma}_k)= \sum_{j=1}^{n_M}\left( \prod_{t'<t, t'\in\widetilde\mathcal{T}_{kj}}[1-h(\boldsymbol{W}_{kj}(t');\boldsymbol{\gamma}_k)] \right)I(C_{kj}\ge t)\boldsymbol{Z}_{kj}(t)\exp(\boldsymbol{\beta}_k^T\boldsymbol{Z}_{kj}(t))$.

\subsubsection{Efron's method for ties}

In observational studies with {\color{black}{age}} in year as time unit, ties commonly exist in the data sets. Our proposed estimating equations (\ref{brewslow_est_eq}) use the Breslow's method to deal with ties \cite{breslow1974covariance}. Efron's tie method may be preferable to the Breslow's method, especially when the sample size is small \cite{efron1977efficiency, hertz1997validity}. In this section, we additionally provide the estimating equations that uses Efron's tie method.

Let $d_k(t)$ represents the number of ties (i.e. total number of potential events across individuals) at time $t$ for the $k$-th outcome, and for $r=1,\ldots,d_{k}(t)$, define
\begin{equation*}
	\begin{split}
		\widetilde{S}_{k}^{(0)}(\boldsymbol{\beta},r,t)&=\sum_i\widetilde{Y}_{ki}(t)\left\{1-\frac{r-1}{d_{k}(t)}h(\boldsymbol{W}_{ki}(t);\boldsymbol{\gamma}_k)\right\}\exp(\boldsymbol{\beta}_k^T\boldsymbol{Z}_{ki}(t))\\
	\widetilde{S}_{k}^{(1)}(\boldsymbol{\beta},r,t)&=\sum_i\widetilde{Y}_{ki}(t)\left\{1-\frac{r-1}{d_{k}(t)}h(\boldsymbol{W}_{ki}(t);\boldsymbol{\gamma}_k)\right\}\exp(\boldsymbol{\beta}_k^T\boldsymbol{Z}_{ki}(t))\boldsymbol{Z}_{ki}(t).\\
\end{split}
\end{equation*} The estimating equations for the $k$-th event become:
\begin{equation}
\begin{split}
	\widetilde{\boldsymbol{U}}_k(\boldsymbol{\beta}_k)&=\sum_{i=1}^{n_M}\int_{0}^{\infty}\frac{1}{d(t)}\sum_{r=1}^{d(t)}\left\{ \boldsymbol{Z}_{ki}(t)- \frac{	\widetilde{S}_{k}^{(1)}(\boldsymbol{\beta},r,t)}{	\widetilde{S}_{k}^{(0)}(\boldsymbol{\beta},r,t)} \right\}\,d\widetilde{N}_{ki}(t)\\
	&=\sum_{i=1}^{n_M}\sum_{t\in\mathcal{T}_{ki}}\left\{\left(  h(\boldsymbol{W}_{ki}(t);\boldsymbol{\gamma_k})\prod_{t'<t, t'\in\widetilde\mathcal{T}_{ki}}  (1-h(\boldsymbol{W}_{ki}(t');\boldsymbol{\gamma}_k)) \right)\frac{1}{d(t)}\sum_{r=1}^{d(t)}\left( \boldsymbol{Z}_{ki}(t)- \frac{	\widetilde{S}_{k}^{(1)}(\boldsymbol{\beta},r,t)}{	\widetilde{S}_{k}^{(0)}(\boldsymbol{\beta},r,t)} \right)\right\}.
\end{split}
\label{efron_est}
\end{equation}

Correspondingly, the equations for estimating the sandwich variance in Equation (\ref{normality}) need to be modified and detailed formula are provided in the supplementary materials Section 1.3.

\subsection{Outcome measurement error-corrected WLW model under the MS/IVS design}

In contrast with the MS/EVS design where the external validation study is collected only for the purpose of building measurement error models, the validation study under MS/IVS design also contributes to the WLW model in addition to building the measurement error model as it is a subset of the main study. We propose two methods to estimate $\boldsymbol{\beta}_k,k=1,\ldots,K$ under the MS/IVS design.

We name the first method below the full calibration method. In the full calibration method, even though the true outcomes are available for participants from the validation study, we treat them as if their true outcomes are missing and, same as the participants not in the validation study, their contributions are also the weighted sum across each potential event time in the estimating equations of the {\color{black}{weighted WLW model}}. {\color{black}{In the supplementary material Section 1.4.1}}, we show that the {\color{black}{estimating functions}} for the measurement error models and the {\color{black}{estimating functions}} for the weighted WLW model are asymptotically independent and therefore, for the full calibration method under MS/IVS design, estimation and inference follow the same steps as in the ME/EVS design. The full calibration method may lose information since the true outcomes of participants in the validation studies are discarded when fitting the weighted WLW model. Next we propose the second method based on a pooled estimator that makes full use of available information.

In the second method, we will fit two separate WLW models. For participants in the validation set, since the true outcomes are available, we fit a standard WLW model with parameters estimated using estimating equations (\ref{WLW_est}); denote the corresponding parameter estimates as $\widehat{\boldsymbol{\beta}}^V$. For participants in the main study without true outcomes available (i.e. excluding participants from the validation study), we fit the proposed weighted WLW model based on either estimating equations (\ref{brewslow_est_eq}) or (\ref{efron_est}), denoting the corresponding parameter estimates as $\widehat{\boldsymbol{\beta}}^M$. The pooled estimator is an inverse variance-weighted average of the two estimates {\color{black}{\cite{spiegelman2001efficient}}}:
\begin{equation*}
\widehat{\boldsymbol{\beta}}^P=({\Sigma}_{\widehat{\boldsymbol{\beta}}^{V}}^{-1}+{\Sigma}_{\widehat{\boldsymbol{\beta}}^{M}}^{-1})^{-1}({\Sigma}_{\widehat{\boldsymbol{\beta}}^{V}}^{-1}\widehat{\boldsymbol{\beta}}^V+{\Sigma}_{\widehat{\boldsymbol{\beta}}^{M}}^{-1}\widehat{\boldsymbol{\beta}}^M),
\end{equation*}
where ${\Sigma}_{\widehat{\boldsymbol{\beta}}^{V}}$ and ${\Sigma}_{\widehat{\boldsymbol{\beta}}^{M}}$ are the variance-covariance matrices for $\widehat{\boldsymbol{\beta}}^{V}$ and $\widehat{\boldsymbol{\beta}}^{M}$, respectively. In practice,  we can plug in the estimated covariance matrices $\widehat{\Sigma}_{\widehat{\boldsymbol{\beta}}^{V}}$ and $\widehat{\Sigma}_{\widehat{\boldsymbol{\beta}}^{M}}$. We refer the readers to Supplementary Material Section 1.4.2 for the estimating equations and technical details.

We can show that $[\widehat{\boldsymbol{\beta}}^M,\widehat{\boldsymbol{\beta}}^V, \widehat{\boldsymbol{\gamma}}]$ jointly follow a multivariate normal distribution:

\begin{equation*}
\begin{split}
	\begin{bmatrix}
		\sqrt{(1-\rho)n_M}(\widehat{\boldsymbol{\beta}}^M-\boldsymbol{\beta})\\
		\sqrt{(1-\rho)n_M}(\widehat{\boldsymbol{\theta}}-\boldsymbol{\theta})
	\end{bmatrix}\overset{\mathcal{L}}{\to}\text{MVN}\left(\boldsymbol{0}, (\boldsymbol{A}^{P}(\boldsymbol{\beta}^M,\boldsymbol{\theta}))^{-1}\boldsymbol{B}^{P}(\boldsymbol{\beta}^M,\boldsymbol{\theta})((\boldsymbol{A}^{P}(\boldsymbol{\beta}^M, \boldsymbol{\theta}))^{-1})^T\right)
\end{split}
\end{equation*}
where $\widehat{\boldsymbol{\theta}}=[\widehat{\boldsymbol{\beta}}^V, \widehat{\boldsymbol{\gamma}}]$, $\boldsymbol{A}^{P}(\boldsymbol{\beta}^M,\boldsymbol{\theta})=\begin{bmatrix}
A^{P}_{\boldsymbol{\beta}^M\boldsymbol{\beta}^{M}}(\boldsymbol{\beta}^M,\boldsymbol{\theta})&A^{P}_{\boldsymbol{\beta}^{M}\boldsymbol{\theta}}(\boldsymbol{\beta}^M,\boldsymbol{\theta})\\
\boldsymbol{0}&A^P_{\boldsymbol{\theta}\boldsymbol{\theta}}(\boldsymbol{\beta}^M,\boldsymbol{\theta})\\
\end{bmatrix}$, and\\ $\boldsymbol{B}^{P}(\boldsymbol{\beta}^M,\boldsymbol{\theta})=\begin{bmatrix}
\boldsymbol{B}^{P}_{\boldsymbol{\beta}^{M}}(\boldsymbol{\beta}^M,\boldsymbol{\theta})&\boldsymbol{0}\\
\boldsymbol{0}&{\color{black}{\frac{(1-\rho)}{\rho}}}\boldsymbol{B}^{P}_{\boldsymbol{\theta}}(\boldsymbol{\beta}^M,\boldsymbol{\theta})
\end{bmatrix}$. Both $\boldsymbol{A}^{P}(\boldsymbol{\beta}^M,\boldsymbol{\theta})$ and $\boldsymbol{B}^{P}(\boldsymbol{\beta}^M,\boldsymbol{\theta})$ can be consistently estimated in a similar way as in the MS/EVS design setting, and detailed formula are provided in the supplementary material Section 1.2. The estimated variance-covariance matrix of the pooled estimator $\widehat{\boldsymbol{\beta}}^P$ can be obtained through the multivariate delta method.

\section{Simulation studies}

We evaluate the finite sample performance of our methods through simulation studies under both the MS/EVS and MS/IVS design. We consider the scenario with $K=2$ events, and self-reported outcomes are collected at 4 questionnaire returns fixed at time 1, 3, 5, and 7. The true event times $T_{1}$ and $T_{2}$ are generated from the Gumbel's bivariate exponential distribution that has the following joint cumulative distribution function \cite{gumbel1960bivariate,wei1989regression}: 
\begin{equation*}
F(t_1,t_2)=F_1(t_1)F_2(t_2)\left[1+\theta\{1-F_1(t_1)\}\{1-F_2(t_2)\}\right].
\end{equation*}
The parameter $\theta,\theta\in (0,1)$ measures the degree of dependence between the two event times with the correlation between $T_{1}$ and $T_{2}$ being $\theta/4$. The corresponding marginal distribution of  $T_1$ and $T_2$ are two univariate exponential distributions with hazard rates $\lambda_1=\exp({\beta}_1Z)$ and $\lambda_2=\exp({\beta}_2Z)$, respectively and $Z$ is a binary exposure variable that is common across two events for a particular individual. To generate the self-reported outcomes, we assume that once the participants report they have experienced the events, their self-reported event statuses would stay the same thereafter, and thus the self-reported outcomes are generated as: 
\begin{equation*}
\delta^\ast_{k}(t)\sim\max\{\delta^\ast_{k}(t-1), \text{Bernoulli}(p^\ast_{k}(t))\}.
\end{equation*}
When the event is yet to be reported, the $k$-th self-reported outcome is generated from a Bernoulli distribution with probability $p^\ast_{k}(t)$, with $\log\frac{p^\ast_{k}(t)}{1-p^\ast_{ki}(t)}=\alpha_0+\alpha_1\delta_{k}(t)$, where $\delta_{k}(t)$ is the true $k$-th event status at each questionnaire return time, and $\alpha_0$, $\alpha_1$ are determined such that different sensitivities and specificities of the self-reported outcomes are achieved. In the simulation, we choose (sensitivity, specificity) among $(0.7,0.7)$, $(0.7,0.9)$, $(0.9,0.7)$ and $(0.9, 0.9)$.

The weighted WLW model fitted in the main study assumes the form: $\lambda_k(t)=\lambda_{k0}(t)\exp(\beta_k Z)$. The measurement error models fitted in the validation study are set to be pooled logistic regressions with true event indicator $\delta_{k}(t)$ as the outcome and the self-reported outcomes $\delta_{k}^\ast(t)$, exposure indicator $Z$ and time $t$ as covariates. We set the sample size of the main study to be 1000 and the validation study to be 50 or 100.

Table 1 and Supplementary tables 1 and 2 contain results under the MS/EVS design, where the baseline hazard rates are set to $\frac{1}{7}, \frac{1}{5}$ and $\frac{1}{9}$, corresponding to an average censoring rate ranging from 19\% to 42\%. The true values for the coefficients $(\beta_1,\beta_2)$ are set among $(\log(1.25),\log(1.5))$, $(\log(1.25),\log(1.75))$ and $(\log(1.5),\log(1.75))$. Percent relative biases of the coefficient estimates and the empirical coverage rates of the 95\% confidence intervals (CI) over the 1000 simulation replicates are reported for both our method and the naive method where the self-reported event statuses are used as the outcomes when fitting the WLW model. We observe heavily biased estimates of the naive method with the relative biases generally being greater than 30\%, regardless of the sensitivities and specificities of the self-reported outcomes, and the corresponding 95\% CI's  have poor coverage rates, which are generally below 80\%. Our method yields point estimates with relative biases generally below 5\% and coverage rates of the 95\% CI's centered around the 95\% nominal level. Moreover, as we increase either the sensitivity, specificity or the size of the validation set, the relative biases of the point estimate decrease.

Table 2 and Supplementary tables 3 and 4 present results under the MS/IVS design. Similar to the results under the MS/EVS design, the naive estimator is still highly biased with poor coverage rates, while our proposed full calibration estimator and the inverse variance pooled estimator both have relative biases less than 5\% and coverage rates centered around the 95\% nominal level in general. {\color{black}{The full calibration estimator and the inverse variance pooled estimator have comparable finite sample performance in terms of both bias and efficiency.}} 

\section{Data Analysis}

We apply our methods to investigate the association between tinnitus and hearing loss in CHEARS. Tinnitus is found to be associated with hearing loss in cross-sectional studies \cite{nicolas2002characteristics}. However, there is little investigation on the temporal relation between tinnitus and hearing loss. One study has shown that participants with persistent tinnitus have higher risk of 3-year hearing threshold elevation in AAA\cite{curhan2021tinnitus}. Since AAA is a subset of CHEARS, focusing on AAA incurs tremendous loss of information. In this section, we aim to conduct an analysis in the entire CHEARS using our proposed methods. The baseline of the analysis is 1991 when the information on study participants' characteristics was available. The self-reported hearing status was collected through the questionnaires distributed in 2009, 2013 and 2017. Whenever a participant reported hearing problems, they also provided the age at which they first noticed the changes in their hearing conditions. Therefore, we treat each year after the baseline as a potential event time for true hearing loss. In AAA, the formal audiometric hearing tests were conducted twice for each study participant, where the first and second tests were conducted in 2012-2015 and 2015-2018, respectively. The hearing loss is defined as the PTA of the better hearing ear being greater than 35 dB at low, mid and high frequencies \cite{wilson2017global}. For participants who were shown to have hearing loss from the baseline testing, {\color{black}{we assume that the true hearing loss time is equal to the time at which they received the formal audiometric hearing test. As a sensitivity analysis, we choose the middle time point between the baseline hearing test time and the starting time (1991) of CHEARS as the event time for people who were detected with hearing loss at baseline testing in AAA. }}

Due to limited number of cases for hearing loss at low frequency, we combine hearing loss at low and mid frequencies and define time to hearing loss at low or mid frequencies as the earliest time of PTA being greater than 35 dB at either low or mid frequency for the better hearing ear. In the analysis, we explore the association between tinnitus (yes/no) on hearing loss at low or mid frequencies and high frequency after adjusting for baseline age, BMI, hypertension (yes/no), smoking status (ever/never), and physical activity. {\color{black}{Table ~\ref{table1} presents a summary of the characteristics of participants grouped by tinnitus status.}} For the measurement error models, we fit two pooled logistic regressions in AAA, one for low or mid frequency and the other for high frequency, with each year as a time interval, starting from baseline of the study. In the pooled logistic regressions, the self-reported hearing status, calendar time, baseline age, BMI, hypertension, tinnitus and physical activity are included as predictors of the true hearing status. Note that, in CHEARS, only overall hearing status was reported by individuals, and therefore for the two event types, they share the same self-reported outcome (i.e. $\delta_{k=1,i}^\ast(t)=\delta_{k=2,i}^\ast(t)$), and can be regarded as a special case of our proposed method.

Table \ref{real} reports the point estimates and 95\% CI's of the effects of tinnitus on hearing loss for low or mid frequency and high frequency from the full calibration method and the inverse variance pooled estimator.  For the naive method that uses the self-reported events as outcomes, since the self-reported outcome only describes the overall hearing ability that are indistinguishable between different frequencies, we use the standard Cox regression. From the naive method there is a significant association of tinnitus with hearing; the hazard ratio is 3.57 (95\% CI: [3.32, 3.84]). The outcome measurement error-corrected results also show significant associations, with alleviated effects. From the full calibration method, the hazard ratios for tinnitus effect on hearing loss at low or mid frequency and high frequency are 2.39 (95\% CI: [1.60, 3.56]) and 2.32 (95\% CI: [1.89, 2.84]), respectively, and from the inverse variance pooled estimator, the hazard ratios are 2.41 (95\% CI: [1.65, 3.52]) and 2.25 (95\% CI: [1.83, 2.76]), respectively. The sensitivity analysis leads to similar point and interval estimates

Moreover, we can perform hypothesis testing on whether the effects of tinnitus on hearing loss are the same across low or mid frequency and high frequency. The hypothesis can be formulated as: $H_0:\beta_{\text{low or mid}}=\beta_{\text{high}} \text{ v.s. }H_1:\beta_{\text{low or mid}}\ne\beta_{\text{high}}$. We take the results from the full calibration method from Table \ref{real} as an example. The Wald test statistic is: $Z=\frac{\widehat{\beta}_{\text{low or mid}}-\widehat{\beta}_{\text{high}}}{\text{SE}(\widehat{\beta}_{\text{low or mid}}-\widehat{\beta}_{\text{high}})}=\frac{\widehat{\beta}_{\text{low or mid}}-\widehat{\beta}_{\text{high}}}{\sqrt{\text{Var}(\widehat{\beta}_{\text{low or mid}})+\text{Var}(\widehat{\beta}_{\text{high}})-2\text{COV}(\widehat{\beta}_{\text{low or mid}},\widehat{\beta}_{\text{high}})}}$. Plugging in the corresponding estimates, the value for the test statistic is $Z=0.16$ (p-value=0.87). Therefore, we do not have sufficient evidence to conclude that tinnitus has heterogeneous effects on hearing loss at different frequencies. 

\section{Discussion}

In this paper, we develop statistical methods to deal with error-prone self-reported outcomes that are collected through questionnaire returns in survival analyses with multiple failure types under the framework of WLW model. Since the true event status is not available, we regard each questionnaire return time as a potential event time that contributes to the estimating equations of the regression parameters, and their contributions are weighted by the probability of the true events occurring at each potential event time. The weights can be estimated by fitting measurement error models using the pooled logistic regressions in the validation study. To account for the uncertainty in the estimation of the parameters in the measurement error models, we derive the sandwich variance estimator based on joint estimating equations for the estimated coefficients. The proposed method can be applied to both the MS/EVS and MS/IVS design. A R program implementing the new methods can be found at: \url{http://www.hsph.harvard.edu/molin-wang/software.} 

The measurement error models are fitted using the pooled logistic regression so that the sequence of $\delta_{ki}(t)$ across the questionnaire return times can be modeled properly. To fit the pooled logistic regression, for convenience, times can be divided into adjacent intervals of equal length. If the self-reported outcomes are collected irregularly, we recommend using the smallest time gap in the data set as the time unit and re-construct the time intervals. We will also consider other risk prediction models in future research to bypass the partition of times. For instance, instead of using the pooled logistic regression, Cox regressions can be fitted in the validation set and the weights at potential event time $t_j$ can be $\widehat{S}_k(t_j)-\widehat{S}_k(t_{j-1})$, where $\widehat{S}_k(t)$ is the estimated survival function from the Cox regression for the $k$-th event. Note that $\widehat{S}_k(t)$ involves the baseline hazard function, which can be estimated using the Breslow estimator. 

{\color{black}{One implicit assumption of our method is that the true events happen at the questionnaire times, while in real practice, this assumption might be violated and interval censoring exists in the data set. In future research, we will extend the method to take into account interval censoring. For instance, Finkelstein (1986) proposed method for proportional hazards model for interval-censored data that maximized the full likelihood to simultaneously estimate the baseline hazards and regression parameters \cite{finkelstein1986proportional} . One method may be to construct our event-specific estimating equations based on the likelihood functions in \cite{finkelstein1986proportional} by adding appropriate weights that reflect the probabilities of true events occurring between each time interval, and use sandwich variance formula to obtain the variance-covariance matrix of the estimated regression parameters.}}

Further, a drawback of defining hearing loss through the better hearing ear is that it classifies participants with unilateral hearing impairment who experience hearing difficulty in only one ear into the healthy hearing category. In future research, under the WLW framework for multiple event types, we will further consider using a mixed effects Cox regression model to analyze hearing  loss for both ears under each frequency to model the correlation between ears \cite{chen2022both, sheng2022analytical}.

\section{Acknowledgements}

{\color{black}{This work is supported by NIH grant R01DC017717.}}

\bibliographystyle{plain}

\bibliography{ref}

\begin{table}[h!]
	\tiny
	\centering  
	\caption{Simulation results under the MS/EVS design. The baseline hazard rate (HR) for both events is set to be $\frac{1}{7}$. The ratio ($\rho$) of the sample size in the validation set to the main study is chosen as 0.05 or 0.1. A total of 1000 simulation replicates are conducted, and the percent bias (\%bias) of the coefficient estimates and coverage rate (CR) of the 95\% confidence intervals are reported for both our proposed outcome measurement error (ME) corrected WLW model and the naive method that uses the self-reported events as outcomes. {\color{black}{The empirical standard errors (SE) of our proposed method is also reported.}}}    
	\begin{tabular}{rrlrrrrrrrr}  \hline\hline               &       &       & \multicolumn{4}{c}{ME corrected WLW} & \multicolumn{4}{c}{Naive WLW} \\   
		\cmidrule(lr){4-7}		 \cmidrule(lr){8-11}
		\multicolumn{1}{l}{$\boldsymbol{\beta}$} &  \multicolumn{1}{l}{$\rho$} & (Sens., Spec.) & \multicolumn{1}{l}{\%bias, $\beta_1$ (SE)} & \multicolumn{1}{l}{\%bias, $\beta_2$ (SE)} & \multicolumn{1}{l}{CR, $\beta_1$} & \multicolumn{1}{l}{CR, $\beta_2$ } & \multicolumn{1}{l}{\%bias, $\beta_1$} & \multicolumn{1}{l}{\%bias, $\beta_2$} & \multicolumn{1}{l}{CR, $\beta_1$} & \multicolumn{1}{l}{CR, $\beta_2$} \\   \midrule
		\multicolumn{1}{l}{$\beta_1=\log(1.25)$} & 0.05  & (0.7, 0.7) & -0.9\% (0.42)& 8.7\% (0.43)& 0.95 & 0.95 & -71.8\% & -69.4\% & 0.33 & 0.01 \\    
		\multicolumn{1}{l}{$\beta_2=\log(1.5)$} &              & (0.7, 0.9) & 2.1\% (0.30)& 5.5\% (0.30) & 0.96 & 0.94 & -38.2\% & -35.9\% & 0.78 & 0.49 \\          
		&       & (0.9, 0.7) & -2.8\% (0.40) & 7.3\% (0.40) & 0.95& 0.95 & -64.1\% & -61.2\% & 0.40 & 0.03 \\          
		&       & (0.9, 0.9) & -0.8\% (0.27) & 4.1\% (0.26)& 0.95& 0.94  & -32.2\% & -29.5\% & 0.83 & 0.62 \\    
		\cmidrule{2-11}
		&  0.1   & (0.7, 0.7) & -0.8\% (0.30) & 4.3\% (0.29) & 0.95 & 0.95& -72.4\% & -70.7\% & 0.32 & 0.01 \\          
		&         & (0.7, 0.9) & 0.4\% (0.22)& 2.7\% (0.21)& 0.95 & 0.94 & -38.9\% & -37.3\% & 0.80 & 0.46 \\          
		&          & (0.9, 0.7) & -0.6\% (0.28)& 2.2\% (0.27)& 0.95 & 0.95 & -64.1\% & -62.6\% & 0.43 & 0.02 \\          
		&        & (0.9, 0.9) & -0.43\% (0.19)& 0.93\% (0.18)& 0.94 & 0.95& -32.5\% & -31.1\% & 0.83 & 0.58 \\    \midrule
		\multicolumn{1}{l}{ $\beta_1=\log(1.25)$ } & 0.05  & (0.7,0.7) & 5.5\% (0.42)& 3.6\% (0.42)& 0.96 & 0.96 & -71.8\% & -69.7\% & 0.35 & 0.00 \\    \multicolumn{1}{l}{ $\beta_2=\log(1.75)$ } &       & (0.7,0.9) & 6.7\% (0.30)& 2.7\% (0.30)& 0.95& 0.95 & -38.0\% & -36.8\% & 0.78 & 0.17\\          &       & (0.9,0.7) & 2.5\% (0.40)& 3.1\% (0.38)& 0.96 & 0.96  & -64.1\% & -61.1\% & 0.41 & 0.00 \\          &       & (0.9,0.9) & 4.4\% (0.27)& 1.4\% (0.25)& 0.96 & 0.95 & -32.2\% & -30.0\% & 0.84 & 0.33 \\     \cmidrule{2-11}      & 0.1   & (0.7,0.7) & 4.7\% (0.30)& 3.8\% (0.28)& 0.95  & 0.94  & -70.5\% & -69.4\% & 0.34 & 0.00\\          &       & (0.7,0.9) & 1.6\% (0.22)& 1.1\% (0.21)& 0.95 & 0.93& -37.7\% & -36.6\% & 0.78 & 0.20 \\          &       & (0.9,0.7) & 3.9\% (0.28)& 3.3\% (0.26)& 0.95 & 0.93& -62.6\% & -60.9\% & 0.41 & 0.00 \\          &       & (0.9,0.9) & 0.1\% (0.19)& 1.1\% (0.18)& 0.96& 0.94& -31.6\% & -29.9\% & 0.82 & 0.33 \\   
		\midrule \multicolumn{1}{l}{$\beta_1=\log(1.5)$} & 0.05  & (0.7, 0.7) & 1.4\% (0.42) & 6.9\% (0.42)& 0.95 & 0.94 & -70.6\% & -69.0\% & 0.01 & 0.00 \\    \multicolumn{1}{l}{$\beta_2=\log(1.75)$} &       & (0.7, 0.9) & 2.2\% (0.30)& 4.2\% (0.30)& 0.95 & 0.95& -37.2\% & -35.7\% & 0.47 & 0.19 \\          &       & (0.9, 0.7) & -0.5\% (0.39)& 6.0\% (0.38)& 0.94 & 0.95 & -62.4\% & -60.2\% & 0.02 & 0.00 \\          &       & (0.9, 0.9) & 0.4\% (0.26) & 3.2\% (0.25)& 0.94 & 0.93& -30.8\% & -28.9\% & 0.58 & 0.37 \\ \cmidrule{2-11}   \textcolor[rgb]{ 1,  0,  0}{} & 0.1   & (0.7, 0.7) & 2.8\% (0.29)& 3.1\% (0.28)& 0.95 & 0.95 & -70.9\% & -69.5\% & 0.01 & 0.00 \\          &       & (0.7, 0.9) & 2.6\% (0.21)& 2.4\% (0.21)& 0.95 & 0.96& -37.6\% & -36.3\% & 0.43& 0.19 \\          &       & (0.9, 0.7) & 2.6\% (0.27)& 1.5\% (0.26)& 0.95 & 0.95  & -62.3\% & -61.0\% & 0.03 & 0.00 \\          &       & (0.9, 0.9) & 1.7\% (0.19)& 0.9\% (0.18)& 0.94& 0.95 & -30.1\% & -29.8\% & 0.58 & 0.33\\    \hline\hline
	\end{tabular}
\end{table}

\begin{sidewaystable}[h!] 
	\tiny 
	\centering  
	\caption{Simulation results under the MS/IVS design. The baseline hazard rate (HR) for both events is set to be $\frac{1}{7}$. The ratio ($\rho$) of the sample size in the validation set to the main study is chosen as 0.05 or 0.1. A total of 1000 simulation replicates are conducted, and the percent bias (\%bias) of the coefficient estimates and coverage rate (CR) of the 95\% confidence intervals are reported for both our proposed outcome measurement error (ME) corrected WLW model and the naive method that uses the self-reported events as outcomes. {\color{black}{Empirical standard errors (SE) of the full calibration method and inverse variance pooled method are also reported.}}}    
	\begin{tabular}{rrlrrrrrrrrrrrr}\hline\hline         
		&       &       & \multicolumn{4}{c}{ME corrected WLW, Full calibration} & \multicolumn{4}{c}{ME corrected WLW, Pooled} & \multicolumn{4}{c}{Naive WLW} \\\cmidrule(lr){4-7}		 \cmidrule(lr){8-11} \cmidrule(lr){12-15}
		\multicolumn{1}{l}{$\boldsymbol{\beta}$} &  \multicolumn{1}{l}{$\rho$} & (Sens., Spec.) & \multicolumn{1}{l}{\%bias, $\beta_1$ (SE)} & \multicolumn{1}{l}{\%bias, $\beta_2$ (SE)} & \multicolumn{1}{l}{CR, $\beta_1$} & \multicolumn{1}{l}{CR, $\beta_2$ }&\multicolumn{1}{l}{\%bias, $\beta_1$ (SE)} & \multicolumn{1}{l}{\%bias, $\beta_2$ (SE)} & \multicolumn{1}{l}{CR, $\beta_1$} & \multicolumn{1}{l}{CR, $\beta_2$ } & \multicolumn{1}{l}{\%bias, $\beta_1$} & \multicolumn{1}{l}{\%bias, $\beta_2$} & \multicolumn{1}{l}{CR, $\beta_1$} & \multicolumn{1}{l}{CR, $\beta_2$} \\   \midrule \multicolumn{1}{l}{$\beta_1=\log(1.25)$} & 0.05  & (0.7, 0.7) & -4.8\%  (0.43)& 7.2\% (0.41)& 0.95 & 0.96 & -5.5\% (0.43)& 5.1\% (0.41)& 0.95 & 0.95 & -72.1\% & -69.3\% & 0.32 & 0.01 \\    \multicolumn{1}{l}{$\beta_2=\log(1.5)$} &       & (0.7, 0.9) & -0.3\% (0.31)& 2.7\% (0.30)& 0.93 & 0.95 & -2.2\% (0.32)& 2.0\% (0.31)& 0.94 & 0.95 & -39.0\% & -36.6\% & 0.77 & 0.47 \\          &       & (0.9, 0.7) & -5.4\% (0.39)& 7.1\% (0.38)& 0.96 & 0.96 & -5.7\% (0.40)& 5.3\% (0.39)& 0.95& 0.96 & -64.6\% & -61.3\% & 0.41 & 0.03 \\          &       & (0.9, 0.9) & 0.5\% (0.26)& 1.2\% (0.25)& 0.95 & 0.94 & -1.0\% (0.27)& 1.2\% (0.27)& 0.95 & 0.95 & -33.0\% & -30.3\% & 0.81  & 0.59\\ \cmidrule{2-15}     \textcolor[rgb]{ 1,  0,  0}{} & 0.1   & (0.7, 0.7) & 2.7\% (0.30)& 5.4\% (0.29)& 0.95 & 0.95 & 0.9\% (0.30)& 4.4\% (0.29)& 0.96 & 0.95 & -72.5\% & -69.7\% & 0.32 & 0.02\\          &       & (0.7, 0.9) & 0.3\% (0.21)& 5.5\% (0.21)& 0.95& 0.95& -1.2\% (0.22)& 4.1\% (0.22)& 0.95 & 0.95 & -37.8\% & -36.4\% & 0.78& 0.46 \\          &       & (0.9, 0.7) & 2.7\% (0.28)& 5.1\% (0.26)& 0.96 & 0.96 & 1.2\% (0.28)& 4.3\% (0.27)& 0.96 & 0.96 & -64.9\% & -61.6\% & 0.41 & 0.043 \\          &       & (0.9, 0.9) & 0.5\% (0.18)& 5.0\% (0.18)& 0.96& 0.96 & -0.8\% (0.19)& 4.0\% (0.19)& 0.95 & 0.96 & -31.7\% & -30.1\% & 0.83 & 0.60 \\  \midrule
		\multicolumn{1}{l}{$\beta_1=\log(1.25)$} & 0.05  & (0.7,0.7) & 3.0\% (0.43)& 8.0\% (0.41)& 0.95 & 0.95 & -5.5\% (0.43)& 3.1\% (0.41)& 0.95 & 0.96& -71.6\% & -69.0\% & 0.32 & 0.000 \\    \multicolumn{1}{l}{$\beta_2=\log(1.75)$} &       & (0.7,0.9) & -0.5\% (0.31)& 6.7\% (0.30)& 0.96& 0.93& -2.2\% (0.32)& 1.4\% (0.31)& 0.94 & 0.95& -39.6\% & -36.5\% & 0.77 & 0.19\\          &       & (0.9,0.7) & 0.7\% (0.39)& 5.9\% (0.38)& 0.96  & 0.95  & -5.7\% (0.40)& 3.4\% (0.38)& 0.95 & 0.96 & -64.1\% & -60.4\% & 0.40& 0.00 \\          &       & (0.9,0.9) & -1.1\% (0.26)& 4.8\% (0.25)& 0.95& 0.94 & -1.1\% (0.27)& 0.8\% (0.26)& 0.95 & 0.96& -33.5\% & -29.8\% & 0.80 & 0.35 \\ \cmidrule{2-15}            & 0.1   & (0.7,0.7) & 4.8\% (0.30)& 0.0\% (0.28)& 0.95 & 0.95& 0.9\% (0.30)& 3.2\% (0.29)& 0.96 & 0.95 & -72.3\% & -69.2\% & 0.32 & 0.00 \\          &       & (0.7,0.9) & 3.9\% (0.21)& 1.2\% (0.21)& 0.94& 0.94 & -1.2\% (0.22)& 3.0\% (0.21)& 0.95 & 0.95& -39.7\% & -36.1\% & 0.76 & 0.19 \\          &       & (0.9,0.7) & 5.0\% (0.27)& -1.0\% (0.26)& 0.94 & 0.95& 1.2\% (0.28)& 3.1\% (0.27)& 0.96 & 0.95 & -64.8\% & -60.6\% & 0.41 & 0.00 \\          &       & (0.9,0.9) & 4.0\% (0.18)& 0.2\% (0.18)& 0.95 & 0.96 & -0.8\% (0.19)& 3.0\% (0.18)& 0.96 & 0.95 & -33.2\% & -29.4\% & 0.81 & 0.33\\  \midrule
		\multicolumn{1}{l}{$\beta_1=\log(1.5)$} & 0.05  & (0.7, 0.7) & -0.8\% (0.42)& 4.8\% (0.41)& 0.95 & 0.95 & -1.5\% (0.42)& 3.1\% (0.41)& 0.95 & 0.96 & -70.5\% & -68.7\% & 0.01  & 0.00 \\    \multicolumn{1}{l}{$\beta_1=\log(1.75)$} &       & (0.7, 0.9) & 1.1\% (0.30)& 2.2\% (0.30)& 0.94& 0.95 & -0.4\% (0.31)& 1.4\% (0.31)& 0.94 & 0.95 & -37.4\% & -36.0\% & 0.45& 0.20 \\          &       & (0.9, 0.7) & -2.0\% (0.38)& 4.9\% (0.38)& 0.95 & 0.95 & -2.1\% (0.39)& 3.4\% (0.38)& 0.95 & 0.96 & -62.5\% & -60.2\% & 0.02 & 0.00 \\          &       & (0.9, 0.9) & 0.7\% (0.25)& 1.1\% (0.25)& 0.95 & 0.94 & -0.2\% (0.27) & 0.8\% (0.26)& 0.95 & 0.96 & -31.2\% & -29.3\% & 0.58 & 0.34 \\\cmidrule{2-15}       \textcolor[rgb]{ 1,  0,  0}{} & 0.1   & (0.7, 0.7) & 2.4\% (0.29)& 4.1\% (0.28)& 0.95 & 0.95& 1.1\% (0.29)& 3.2\% (0.29)& 0.95& 0.95 & -71.0\% & -68.9\% & 0.01 & 0.00 \\          &       & (0.7, 0.9) & 1.0\% (0.21)& 4.1\% (0.21)& 0.95 & 0.95& -0.1\% (0.22)& 3.0\% (0.21)& 0.95 & 0.95 & -36.8\% & -35.8\% & 0.48 & 0.20 \\          &       & (0.9, 0.7) & 2.4\% (0.26)& 3.8\% (0.26)& 0.96& 0.96 & 1.2\% (0.27)& 3.0\% (0.27)& 0.96 & 0.95 & -62.8\% & -60.3\% & 0.03 & 0.00 \\          &       & (0.9, 0.9) & 1.1\% (0.18)& 3.8\% (0.18)& 0.95  & 0.96 & 0.2\% (0.19)& 3.0\% (0.18)& 0.95 & 0.95 & -30.5\% & -29.2\% & 0.59 & 0.34 \\   
		\hline\hline
	\end{tabular}  \label{tab:addlabel}\end{sidewaystable}%

\begin{table}[htbp]  \footnotesize\centering  \caption{  {\color{black}{Summary statistics of baseline characteristics for participants in CHEARS}} }    \begin{tabular}{llll}    \toprule    \multirow{2}[2]{*}{} & \textbf{No Tinnitus} & \textbf{Tinnitus} & \textbf{Total} \\          & \textbf{(N=67601)} & \textbf{(N=6719)} & \textbf{(N=74320)} \\    \midrule    \textbf{Baseline age} &       &       &  \\    Mean (SD) & 36.1 (4.65) & 37.4 (4.49) & 36.2 (4.65) \\    Median [Min, Max] & 36.0 [26.0, 46.0] & 38.0 [26.0, 45.0] & 36.0 [26.0, 46.0] \\    \textbf{Smoking status} &       &       &  \\    never & 45239 (66.9\%) & 4374 (65.1\%) & 49613 (66.8\%) \\    ever  & 22362 (33.1\%) & 2345 (34.9\%) & 24707 (33.2\%) \\    \textbf{BMI} &       &       &  \\    $<25$   & 45440 (67.2\%) & 4230 (63.0\%) & 49670 (66.8\%) \\    25-29 & 13585 (20.1\%) & 1443 (21.5\%) & 15028 (20.2\%) \\    30-34 & 5173 (7.7\%) & 627 (9.3\%) & 5800 (7.8\%) \\    35-39 & 2166 (3.2\%) & 271 (4.0\%) & 2437 (3.3\%) \\    40+   & 1237 (1.8\%) & 148 (2.2\%) & 1385 (1.9\%) \\    \textbf{Hypertension} &       &       &  \\    No Hypertension & 63563 (94.0\%) & 6211 (92.4\%) & 69774 (93.9\%) \\    Hypertension & 4038 (6.0\%) & 508 (7.6\%) & 4546 (6.1\%) \\    \textbf{Diabetes} &       &       &  \\    No Diabetes & 67076 (99.2\%) & 6658 (99.1\%) & 73734 (99.2\%) \\    Diabetes & 525 (0.8\%) & 61 (0.9\%) & 586 (0.8\%) \\    \textbf{Physical Activity, METS, in quintiles} &       &       &  \\    Quntile 1 & 12951 (19.2\%) & 1384 (20.6\%) & 14335 (19.3\%) \\    Quntile 2 & 13719 (20.3\%) & 1443 (21.5\%) & 15162 (20.4\%) \\    Quntile 3 & 13585 (20.1\%) & 1402 (20.9\%) & 14987 (20.2\%) \\    Quntile 4 & 13802 (20.4\%) & 1275 (19.0\%) & 15077 (20.3\%) \\    Quntile 5 & 13544 (20.0\%) & 1215 (18.1\%) & 14759 (19.9\%) \\    \bottomrule    \end{tabular}  \label{table1}
\end{table}

\begin{table}[h!]  
	\centering  
	\caption{The effect of having tinnitus on hearing loss at low or mid frequencies and high frequency in CHEARS. The true hearing loss time is assumed to be equal to the baseline testing time if participants were detected with hearing loss at baseline testing; in the sensitivity analysis the middle time point between the baseline hearing test time and the starting time of CHEARS is used as the event time. }  
	\begin{tabular}{cccc}     \hline\hline \multicolumn{1}{l}{Method} & Frequency & \multicolumn{1}{l}{Tinnitus effect} &\multicolumn{1}{l}{{\color{black}{Sensitivity analysis}}} \\    \midrule    \multicolumn{1}{l}{Naive method} &    -   &3.57 (3.32, 3.84)  & - \\   
		\midrule \multicolumn{1}{l}{Full calibration} & Low or mid &2.39  (1.60, 3.56) & 2.52 (1.73, 3.66)\\          
		& High  & 2.32  (1.89, 2.84) & 2.41 (1.98, 2.94)\\  
		\midrule
		\multicolumn{1}{l}{Pooled estimator} & Low or mid &2.41 (1.65, 3.52)& 2.46 (1.71, 3.55)  \\          
		& High  &  2.25 (1.83, 2.76)&  2.28 (1.88, 2.77) \\    \hline\hline
	\end{tabular} 
	\label{real}
\end{table}%

\end{document}